\documentclass[aps,twocolumn,showpacs,floatfix]{revtex4}

\usepackage{dcolumn,multirow}
\usepackage{graphicx}
\usepackage{amsfonts,mathtools,amsmath,bm,amssymb}
\usepackage{comment}
\usepackage[usenames]{color}
\usepackage[flushleft]{threeparttable}
\usepackage{titlesec}
\usepackage{etoolbox} 
\usepackage{lipsum} 
\usepackage[capitalize]{cleveref}
\usepackage{makecell}

\usepackage{color}

\begin{document}

\title{Origin of the orbital polarization of Co$^{2+}$ in La$_2$CoTiO$_6$ and (LaCoO$_3$)$_1$+(LaTiO$_3$)$_1$ : \\
 a DFT+$U$ and DMFT study }

\author{Alex Taekyung Lee}
\affiliation{Department of Applied Physics, Yale University }

\author{Hyowon Park}
\affiliation{Department of Physics, University of Illinois at Chicago}

\author{Sohrab Ismail-Beigi}
\affiliation{Department of Applied Physics, Yale University}

\date{\today }

\begin{abstract}
The unequal electronic occupation of localized orbitals (orbital polarization), and associated lowering of symmetry and degeneracy, play an important role in the properties of transition metal oxides.  Here, we examine systematically the underlying origin of orbital polarization, taking as exemplar the 3$d$ manifold of Co$^{2+}$ in a variety of spin, orbital and structural phases in the double perovskite La$_2$CoTiO$_6$ and the (001) superlattice (LaCoO$_3$)$_1$+(LaTiO$_3$)$_1$ systems.  Superlattices are of specific interest due to the large experimentally observed orbital polarization of their Co cations.  Based on first principles calculations, we find that robust and observable orbital polarization requires symmetry reduction through the lattice structure; the role of local electronic interactions is to greatly enhance the orbital polarization.

\end{abstract}


\maketitle

\section{Introduction}

The intriguing electronic and magnetic properties of transition metal oxides (TMOs) are governed by the electronic states dervied from their $d$ orbitals.  The associated energy bands lie near the Fermi level, and  the electronic and magnetic properties of TMOs are strongly dependent on the symmetry and degeneracy of the active $d$ orbitals.
High-temperature superconductivity in cuprates \cite{LeeRMP2006,DagottoRMP1994,RMP2003,ZhangPRB1988}, 
phase transitions in manganites \cite{KonishiJPSJ1999,Tokura2000,PRBNanda2010,PesqueraNComm},
metal-insulator transitions in titanates, vanadates, and double perovskties  \cite{ImadaRMP1998,HaverkortPRL2005,alex-dp},
and spin-state transitions in cobaltates \cite{ZobelPRB2002,PRB2020Hyowon}
provide classic examples.  The degree of broken orbital degeneracy and resulting net differences in orbital populations on the transition metal sites across the unit cell of the material, termed ``orbital polarization'', is an important ingredient and the focus on this work.  

Understanding the properties of TMOs is complicated because the charge, orbital, spin, and lattice degrees of freedom are strongly coupled \cite{Tokura2000,Hwang2012} and the physical properties of and phase transitions in TMOs derive from a combination of these couplings.  Orbital polarization can be induced by electron-electron ($e$-$e$) or electron-lattice ($e$-$l$) couplings separately or by mixture of the two.  Disentangling the $e$-$e$ and $e$-$l$ effects on the orbital properties in real materials is challenging, since both mechanisms result in orbital polarization/ordering and concomitant lattice distortions. In addition, there are many modes for $e$-$l$ coupling in TMOs
such as local Jahn-Teller (JT) distortions or  oxygen octahedral tilts and rotations.

Describing the connection between local orbital occupations and orders, local atomic-scale structure, and electron-electron interactions has a rich and long history often associated with classic textbook names (e.g., Jahn-Teller, Goodenough-Kanamori, Kugel'-Khomskii, etc.).  
In terms of particular examples, the metal-insulator transition of Fe$_3$O$_4$ $\sim$120K is explained by a symmetry-lowering charge ordering 
with concomitant structural phase transition \cite{WrightPRB2002,SchlappaPRL2008,AnisimovPRL2004}. 
Additionally, the material shows transitions between ferroelectric and paraelectric phases.  The charge order and polarization connect directly to local structural perturbations, and the relation of local structure to electronic and magnetic properties have been elucidated previously
\cite{YamaguchiPRB2009,SennNature2012}.
Another example involves the orbital  polarization of La$_{1-x}$Sr$_x$MnO$_3$ manganites which has been studied extensively 
\cite{Tokura2000,PRBNanda2010,TebanoPRL2008,TebanoPRB2010,PesqueraNComm} in terms of the strength of the Jahn-Teller distortion and the hole doping level ($x$).

However, how one should create orbital polarization from a {\em materials engineering} viewpoint is not clearly addressed in the literature.  What  material structure is necessary to create strong orbital polarization?  Should one focus primarily on lattice symmetry lowering or strong electronic interactions, and are electronic interactions alone sufficient to  create strong orbital polarization spontaneously?

Disentangling the various possibilities is not trivial since prior work focuses on materials in their most stable ground state structures where all the factors act simultaneously.  For example, in our prior work, we have engineered and characterized systems with strong orbital polarization in nickelate and cobaltate superlattices~\cite{ChenPRL2013,DisaPRL2015,JaePRL2019}.  However, 
since the superlattice geometry automatically breaks structural symmetry, 
the separate effects and relative importance of (i) structure, (ii) symmetry reduction, and (iii) electronic interactions is unclear.  As a typical example, a Jahn-Teller (JT) distortion in a bulk material breaks orbital degeneracy, but it distorts the structure and reduces symmetry at the same time.  In contrast, if one forms an oxide superlattice but enforces the local atomic-scale structure to have perfect cubic symmetry, the system already lowers its symmetry from cubic to tetragonal even before any JT distortion appears.  

Here, we focus on Co cations in TMOs as exemplar systems where strong orbital polarization can be engineering and observed.  Co cations can have multiple spin states, and separately can have active (open-shell) $t_{2g}$ or $e_g$ orbitals depending on their spin state and valence.  Bulk LaCoO$_3$ (LCO) containing Co$^{3+}$ is well known for having multiple spin states: it is a low-spin (LS) state ($t^{6}_{2g}$, $S$=0) nonmagnetic insulator at low temperatures \cite{BhidePRB1972,ZobelPRB2002}, a
paramagnetic insulator for temperatures between 100 and 500 K with either a high-spin (HS) state ($t^{4}_{2g}e^{2}_{g}$, $S=2$) \cite{AsaiPRB1994,ItohJPSJ1994}
or an intermediate spin (IS) state ($t^{5}_{2g}e^{1}_{g}$) \cite{PotzePRB1995,SaitohPRB1997,AsaiJPSJ1998,YamaguchiPRB1997,ZobelPRB2002}, and is metallic above 500 K.  However, the orbital polarization of both the HS and LS states of Co$^{3+}$ is zero in LaCoO$_3$ due to its high symmetry.  
Recently, we have found remarkably strong orbital polarization of Co$^{2+}$ 
in LaCoO$_3$+LaTiO$_3$ (LCO+LTO) superlattices \cite{JaePRL2019}.
Similar to Co$^{3+}$, Co$^{2+}$ has multiple spin states, but is missing the IS state so only HS ($t^{5}_{2g}e^{2}_{g}$) and LS ($t^{6}_{2g}e^{1}_{g}$) are relevant.  While the orbital polarization is mainly due to the \emph{minority spin} $t_{2g}$ orbitals for the HS state,
the polarization for the LS state is due to the \emph{majority spin} $e_g$ bands.  Therefore, this material provides a single system where multiple types of orbital polarization can be studied.  We note that strong orbital polarization can also be engineered in nickelate superlattices in a similar fashion \cite{ChenPRL2013,DisaPRL2015}.

In this work, we use first principles electronic structure calculations based on DFT+$U$ theory \cite{LDA+U1} as well as dynamical mean field theory (DMFT) \cite{PRB2020Hyowon} to study these TMO systems.  We elucidate the origin of orbital polarization in both $e_g$ and $t_{2g}$ manifolds and disentangle the role  of $e$-$e$ and $e$-$l$ couplings. 

We note that the related but distinct term ``orbital ordering'' also refers to unequal orbital populations arranged in a repeating pattern in the unit cell, but observed orbital orderings in TMOs generally have alternating or staggered orbital populations between the different transition metal sites in a single unit cell leading to much smaller net orbital polarization.  In this work, we are focus on orbital polarization itself: the orbital occupations of the Co$^{2+}$ sites in each unit cell are identical  (or extremely similar), making for a cleaner analysis.

\section{Computational Details}
\subsection{Structures and DFT+$U$}
We use density functional theory (DFT) with the projector augmented wave (PAW) method \cite{PAW} and the revised version of the generalized gradient
approximation (GGA) proposed by Perdew {\it et al.} (PBEsol) \cite{PBEsol} as implemented in the VASP software \cite{VASP}.
In all cases, the spin-dependent version of the exchange correlation functional is employed.
A plane wave basis with a kinetic energy cutoff of 500 eV is used. We study the (001)  (LaCoO$_3$)$_1$+(LaTiO$_3$)$_1$ superlattice, denoted as LCO+LTO below.  For the high symmetry structure with $a^0a^0a^0$ octahedral tilts (i.e., no oxygen octahedron tilts or rotations), corresponding to the  $Fm\bar{3}m$ and $P_{4}/mmm$ space groups,  we used 10 atom unit  cells (i.e., a $(1\times1)$ interfacial unit cell). We used 20 atom unit cells (i.e., $c(2\times 2)$ interfacial unit cells) for the $a^-a^-b^+$ tilt structure which has the $P2_{1}/n$ space group. We use $\Gamma$-centered \textbf{k}-point meshes of size  9$\times$9$\times$9 ($Fm\bar{3}m$) and 13$\times$13$\times$7 ($P_{4}/mmm$)
for the 10 atom cells, and  9$\times$9$\times$7 for the 20 atom  cells. 
For more precise calculations of the energy differences listed in Table~\ref{energytable}, we used a kinetic energy cutoff of 700 eV and 17$\times$17$\times$17  \textbf{k}-point meshes.  The GGA+$U$ scheme within the rotationally invariant formalism together with the fully localized
limit double-counting formula \cite{LDA+U1} is used to study the effect of electron interactions.  

Atomic positions within the unit cells were relaxed until the residual forces were less than 0.01 eV/\AA.
For cases with reduced symmetry, the stress was relaxed only along the $z$ axis to be below 0.02 kB, while the in-plane lattice parameters $a$ and $b$ were set equal and took the values 3.811, 3.851, or 3.891 \AA\ in order to simulate the realistic experimental situation where the superlattice is grown as an epitaxial thin film on a substrate.  For the double perovskite La$_2$CoTiO$_6$, we used a face-centered cubic unit cell containing 10 atoms, 
and the lattice parameters correspond to 3.891 \AA.
We note that 3.891 \AA\  is obtained from by minimizing all stresses with $U_{\textnormal{Co}}=U_{\textnormal{Co}}=3$ eV.
We consider both ferromagnetic and antiferromagnetic spin orders: however, for simplicity, we focus on the ferromagnetic case unless specified.

The electronic and structural properties critically depend on the
$U_{\textnormal{Co}}$ value used for the Co 3$d$ manifold, and  we  explore a range of values.  We also explore how the results depend on $U_{\textnormal{Ti}}$, 
which plays a secondary but still important role in the physics of these materials.
We do not employ an on-site exchange interaction $J$ for any species, as the exchange interaction is already accounted for within the spin-dependent DFT exchange-correlation potential \cite{Hyowon,Chen2015}.    

Since the spin-orbit interactions for 3$d$ transition metal atoms are weak, we do not include spin-orbit coupling (SOC) in our calculations and expect their inclusion to lead to only small quantitative changes.  
Explicit testing shows only $\approx$1\% difference between orbital occupancies from GGA+$U$ vs. GGA+$U$+SOC (see Appendix A \ref{appendix:soc}). 

Projected density of states are obtained by the spherical harmonic projections inside spheres around each atom.
Wigner-Seitz radii of 1.323 and 1.302 \AA~ were used for the projection of Ti and Co atoms, 
respectively, as implemented in the VASP-PAW pseudopotential.  
Density matrices for the Co $3d$ manifold are computed using projector functions based on the PAW methodology as implemented in VASP,
following existing frameworks \cite{PRB2000Bengone,JPCM2003Rohrbach}.
Core radii for projector operators of Ti and Co are 1.357 and 1.249 \AA, respectively.

\subsection{DFT+DMFT}

We employ the non-charge-self-consistent DFT+DMFT method \cite{PRB2020Hyowon} for structures obtained from DFT+$U$ calculations for La$_2$TiCoO$_6$ and LTO+LCO with lattice parameter 3.891 \AA. We solve the many-body problem only on the manifold of Co 3$d$ Wannier orbitals that describe the Co-derived conduction bands: physically, these are the states that show broken orbital symmetry. 
The DFT+DMFT calculation has the following steps.
First, we solve the non-spin-polarized Kohn-Sham (KS) equation within DFT+$U$ using the VASP code.
Second, we construct the Hamiltonian of Co 3$d$ bands using maximally localized Wannier functions (MLWFs) \cite{MLWF}.
In the first and the second step, we use $U_{\rm Ti}$=8 eV and $U_{\rm La}$=5 eV, 
to ionize the Ti occupation and minimize the La $d$-Co $d$ hybridization, respectively. 
In the last step, we solve the DMFT self-consistent equations of the correlated subspace of Co 3$d$ Wannier orbitals,
using the continuous time quantum Monte Carlo (CTQMC) \cite{PRBHaule2007,ctqmcRMP2011}
impurity solver to solve the DMFT self-consistent equations. 

Both Hubbard $U$ and Hund's couplings $J$ are parameterized by Slater integrals ($F^0$, $F^2$, and $F^4$), 
using $U = F^0$ and $J = (F^2 + F^4)/14$. 
The Coulomb interaction matrix elements with only density-density types are considered in the CTQMC 
while the spin-flip and pair-hopping terms are neglected. 
We note that the cartesian axes of the Wannier orbitals and the directions of the Co-O bonds are found to be parallel,
so that the off-diagonal terms in the $d$ Hamiltonian are negligibly small.
Therefore, the off-diagonal terms in the DMFT hybridization function are neglected during the CTQMC calculation.

For the DMFT calculations, we used $U$ values of 3 and 6 eV and $J$ of 0.5 and 0.9 eV.
Here we focus on $U$=3 eV, which provides a bulk band gap $\sim$ 1 eV for LaCoO$_3$ as per prior DMFT work \cite{PRLKarolak2015}.
The two values of Hund's coupling $J$ allow us to obtain the low-spin and high-spin states of Co$^{2+}$ (by tuning the relative strength of the Hund's coupling relative to the crystal field).
Specifically, the low-spin state is stable when $J=0.5$ eV, and high-spin state is obtained  when $J=0.9$ eV.
The $J$-dependence of the Co$^{2+}$ spin state is similar to that of Co$^{3+}$ in bulk LaCoO$_3$ \cite{PRLKarolak2015}.
We used electronic temperatures of 150 K and 300 K to study the temperature effect on the spectral function.
Since the results are qualitatively very similar, 300 K results will be discussed unless specified otherwise.

\section{ Orbital polarization of Co$^{2+}$}

\subsection{LS and HS states: basics}
\label{sec:spinstates}

\begin{figure}
\begin{center}
\includegraphics[width=0.5\textwidth, angle=0]{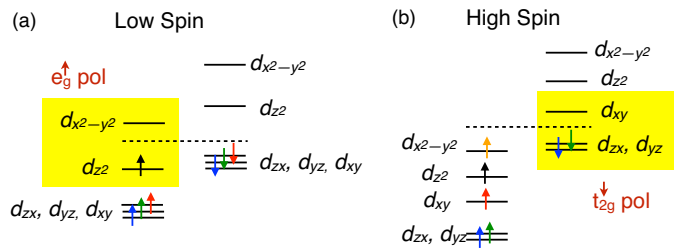}
\caption{  Schematics energy levels of Co$^{2+}$ ions in (LaCoO$_3$)$_1$+(LaTiO$_3$)$_1$ superlattices with majority spin up for (a) low-spin state with $e_g$ orbital polarization, and (b) high-spin state with $t_{2g}$ orbital polarization. This schematic picture holds for both $P_{4}/mmm$ and $P2_1/n$ phases. The yellow shaded states are those determining the orbital polarization in each case.
}
\label{spin_states}
\end{center}
\end{figure}

We begin with a discussion of the basic electronic and magnetic properties of LCO+LTO superlattices 
which is summarized in our previous studies \cite{JaePRL2019,part1}.
In LCO+LTO superlattices as well as the double perovskite La$_2$CoTiO$_6$, there is a charge transfer from Ti$^{3+}(d^1)$  to Co$^{3+}(d^6)$, resulting in 
Ti$^{4+}(d^0)$ and Co$^{2+}(d^7)$. 
The role of electron transfer between Ti and Co has been discussed in prior work~\cite{JaePRL2019}.

We now highlight some basic facts about the Co$^{2+}$ spin states in the systems studied here.  Since the electronic structure of Co$^{2+}$ is strongly dependent on the crystal structure, in this  subsection we will focus only on the $P2_1/n$ phase of the superlattice, which is the most stable phase we have found \cite{JaePRL2019,part1}.

The LS state ($t^{6}_{2g}e^{1}_{g}$) has $S=1/2$ and is illustrated by Fig.~\ref{spin_states}(a):
the $t_{2g}$ states are fully occupied, while the one remaining electron is in the $e_g$ channel.  
In the superlattice, the degeneracy of the $e_g$ manifold is already broken at $U_{\textnormal{Co}}=0$
due to  interface formation and epitaxial strain, with a lower energy $d_{z^2}$ band and higher energy $d_{x^2-y^2}$  band.
When $U_{\textnormal{Co}}\geq1.5$ eV, the $e_g$ bands completely split in energy, resulting in an insulating phase:
only the spin-up $d_{z^2}$ is filled in the LS state while the spin-up $d_{x^2-y^2}$ and spin-down $e_g$ bands are empty
(see the Appendix \ref{appendix:pdos} for associated densities of states).
As a result, the LS state has strong $e_g$ orbital polarization:
we find that the polarization is nonzero at $U_{\textnormal{Co}}=0$, 
and $U_{\textnormal{Co}}>0$ simply enhances it.

The HS state  ($t^{5}_{2g}e^{2}_{g}$) with $S=1$ is depicted in Fig.~\ref{spin_states}(b):
the spin-up $d$ bands are fully occupied, while spin-down $d$ bands have two electrons in the $t_{2g}$ channel.
Unlike LS state, the HS state is not even metastable if $U_{\textnormal{Co}}<1$ eV.
When $U_{\textnormal{Co}}=1$ eV, the $t_{2g}$ bands split into two nearly-degenerate bands ($d_{xz}$ and $d_{yz}$)
and a single $d_{xy}$ band (see the Appendix \ref{appendix:pdos} for the relevant densities of states.). 
We note that $d_{xz}$ and $d_{yz}$ are degenerate for the tetragonal phase ($P_{4}/mmm$),
but this degeneracy is broken in the monoclinic phase ($P2_1/n$).
For $1.5 \leq U_{\textnormal{Co}} <2.5$ eV, the $d_{xy}$ band is completely split in energy from the $d_{xz}$/$d_{yz}$ bands.
However, the spin-down $d_{xy}$ band is partially occupied and the spin-up $d_{x^2-y^2}$ bands is partially empty,
thus the system remains metallic.
When $U_{\textnormal{Co}}\geq 2.5$ eV, the spin-down $d_{xy}$ becomes empty, and the spin-up $d_{x^2-y^2}$ band is fully occupied,
resulting in an insulating phase
(see the Appendix \ref{appendix:pdos} for relevant densities of states).

\subsection{Structural phases}

\begin{figure}
\begin{center}
\includegraphics[width=0.42\textwidth, angle=0]{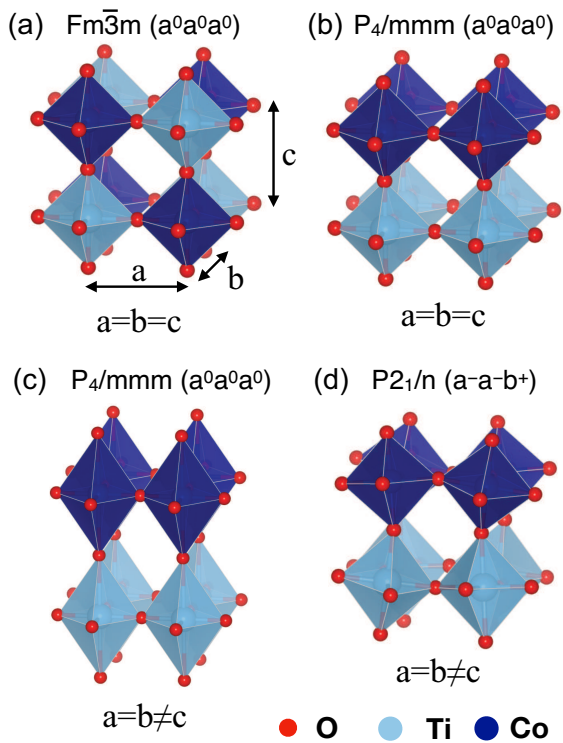}
\caption{  Schematics of the atomic-scale structures of structures studied in this work.  Each structure is labeled by its space group and octahedral rotation pattern. 
 (a) ideal La$_2$CoTiO$_6$ double perovskite, (b) (c) and (d) shown one repeat of LCO+LTO (001) superlattices where the (001) direction is vertical.
La atoms are not shown for clarity. 
 }
\label{str_sym}
\end{center}
\end{figure}

The $e_g$ polarization of the LS state and the $t_{2g}$ polarization of the HS state can be 
due to $e$-$e$ and/or $e$-$l$ coupling.
To disentangle the effect of these two interactions on the orbital polarization,
we consider and compare several reference lattice structures as presented in Fig.~\ref{str_sym}.
(a) We start with the ideal double perovskite La$_2$CoTiO$_6$,
which has the $Fm\bar{3}m$ space group and no octahedral distortions (Fig. \ref{str_sym}(a)) and
where $a=b=c$ and the atomic positions are frozen at ideal cubic perovskite coordinates.
(b) Next, we have a (LaCoO$_3$)$_1$+(LaTiO$_3$)$_1$ superlattice obtained by swapping half the Ti and Co in the ideal double perovskite to create a layered superlattice while keeping idealized atomic coordinates and lattice parameters (this has the $P_{4}/mmm$ space group; see Fig. \ref{str_sym}(b)).
(c) Another $P_{4}/mmm$ phase superlattice where only the atomic positions and the stress along the $c$ axis are relaxed (see Fig. \ref{str_sym}(c)).  
(d) Finally, a $P2_{1}/n$ phase of the LCO+LTO superlattice which has the $a^-a^-b^+$ type of  octahedral tilt \cite{Woodward1}
and is the ground state of the superlattice.
The $P2_{1}/n$ phase is monoclinic, but since we have assumed the epitaxial strain condition where $a=b \neq c$, this differs from a generic monoclinic structure where $a\neq b \neq c$.

\subsection{LS state: $e_g$ orbital polarization}

In this subsection, we focus on the low-spin (LS) state which can have $e_g$ orbital polarization.  

We note that we find zero or insignificant orbital ordering and associated robust JT distortions for in-plane Co--O bonds
in any of the LS phases we  consider:
for the $Fm\bar{3}m$  and $P_{4}/mmm$ phases the high degree of symmetry precludes it, 
while for the $P2_{1}/n$ phase we find negligible  ($<0.002$ \AA) such distortions.
Separately, out-of-plane Co--O bonds can have JT distortion in the $P_{4}/mmm$ and $P2_{1}/n$ phases 
but this does not give rise to orbital ordering either.

\begin{figure}
\begin{center}
\includegraphics[width=0.48\textwidth, angle=0]{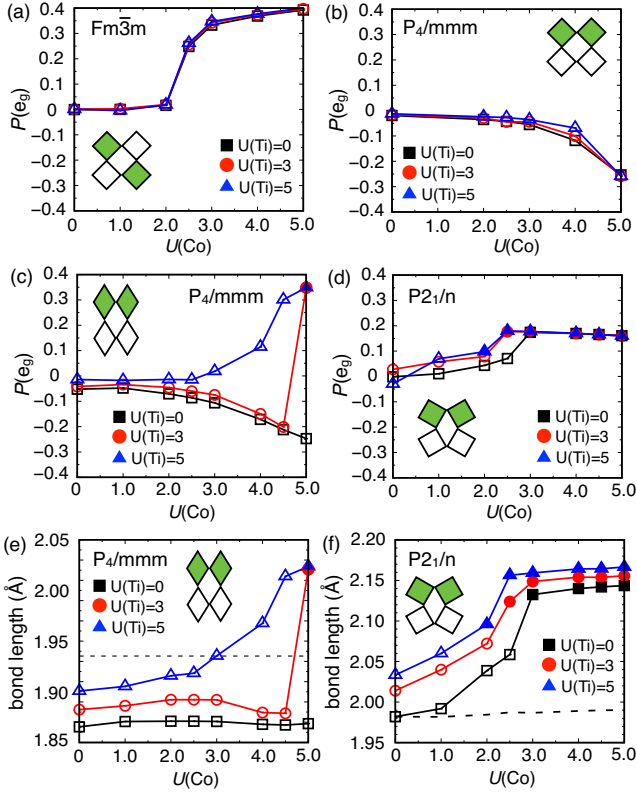}
\caption{(a)-(d) show the DFT+$U$ orbital polarization $P(e_g)$ of the LS state structures with different space groups versus $U({\rm Co})$ and as a function of $U({\rm Ti})$.
Empty and filled points indicate metallic and insulating phases, respectively.
Insets show schematic side views of the octahedral tilts and distortions of the CoO$_6$ and TiO$_6$ oxygen octahedra. 
(e) and (f) present Co--O bond lengths along the $c$ axis of the  $P_{4}/mmm$ and $P2_{1}/n$ phases, respectively.
Dashed lines represent in-plane Co--O bond lengths which depend weakly on $U$.
}
\label{orbpol_eg}
\end{center}
\end{figure}

We define the orbital polarization of the LS state as 
\begin{equation}
P(e_{g}) = \frac{(n^{\uparrow}_{z^2}+n^{\downarrow}_{z^2})- (n^{\uparrow}_{x^2-y^2}+n^{\downarrow}_{x^2-y^2})}
{(n^{\uparrow}_{z^2}+n^{\downarrow}_{z^2})+ (n^{\uparrow}_{x^2-y^2}+n^{\downarrow}_{x^2-y^2})}
\label{eq:eg}
\end{equation}
where the occupancy $n^{\sigma}_i$ is the electron population of orbital $i$ with spin $\sigma$ which is found on the diagonal elements of the single particle density matrix in the Co 3$d$ manifold. 
Figs. \ref{orbpol_eg}(a)-(d) present the DFT+$U$-calculated orbital polarization $P(e_g)$ for the four different structural phases in the LS state as a function of $U_{\textnormal{Co}}$ and $U_{\textnormal{Ti}}$.

\begin{table}
\begin{ruledtabular}
\begin{center}
\caption{Energy difference (in meV/Co) from DFT+$U$ between different Co configurations in the same structure with $U_{\textnormal{Ti}}=3$ eV and $U_{\textnormal{Co}}=5$ eV.  The configurations are written assuming majority up spin electrons, and only the occupancy of the orbitals of interest are shown.  E.g., for the LS $e_g$ case, the full configuration corresponding to the nomenclature $d_{z^2,\uparrow}^1$ is $d_{t_{2g}}^6 d_{z^2,\uparrow}^1$.
For $Fm\bar{3}m$,  $d_{xy,\downarrow}^1(d_{xz,\downarrow}/d_{yz,\downarrow})^1$
means that either the $d_{xz,\downarrow}$ or $d_{yz,\downarrow}$ is filled for all Co cations.
For $P_4/mmm$,  $d_{xy,\downarrow}^1(d_{xz,\downarrow}/d_{yz,\downarrow})^1$
means checkerboard orbital ordering and alternating $d_{xz,\downarrow}^1$ and $d_{yz,\downarrow}^1$ Co  occupations.
}
\label{energytable}
\renewcommand{\arraystretch}{1.3}
\begin{tabular}{c c | c || c | c}
Structure & \multicolumn{2}{c||}{LS $e_g$}	& \multicolumn{2}{c}{HS $t_{2g}$} \\
\cline{2-3} \cline{4-5}
	& $d_{z^2,\uparrow}^1$	& $d_{x^2-y^2,\uparrow}^1$	& $d_{xz,\downarrow}^1d_{yz,\downarrow}^1$		& $d_{xy,\downarrow}^1(d_{xz,\downarrow}/d_{yz,\downarrow})^1$ \\
 \hline
(i) $Fm\bar{3}m$			& 0	& 0.7			& 0	& 0.1 \\	 
\hline
(ii) \makecell{$P_{4}/mmm$ \\ ($a=b=c$)} 		& 0	&  $-$30		& 0	& 66 \\	
\hline
(iii) \makecell{$P_{4}/mmm$ \\ ($a=b\neq c$)} 	& 0	& 21			& 0	& 150 \\	
\end{tabular}
\end{center}
\end{ruledtabular}
\end{table}

\subsubsection{LS $Fm\bar{3}m$ state: lack of orbital polarization}
We begin our analysis with the $Fm\bar{3}m$ space group La$_2$CoTiO$_6$ double perovskite structure (Figs. \ref{str_sym}(a) and \ref{orbpol_eg}(a)).  While $P$ for the $Fm\bar{3}m$ is zero for $U_{\textnormal{Co}}\leq 1$ eV, it becomes significant for $U_{\textnormal{Co}}\geq 2$ eV.  This happens because of spontaneous electronic symmetry breaking: for large enough $U_{\textnormal{Co}}$, the DFT+$U$ total energy is lowered by having the $e_g$ electron occupy one of the two $e_g$ orbitals more than the other.  However, $P\ne0$ for $Fm\bar{3}m$ does not necessarily indicate an actual nonzero orbital polarization in the true interacting system because a single-determinant DFT+$U$ description cannot capture the fluctuations between the $d_{z^2,\uparrow}^1$ and $d_{x^2+y^2,\uparrow}^1$ configurations.  But, the total energies of the two separate configurations should be well captured by DFT+$U$.  Table ~\ref{energytable} shows that these two configurations are essentially degenerate in energy for $Fm\bar{3}m$ (a fully converged DFT+$U$ calculation should find them exactly degenerate): the degeneracy means that we should expect fluctuations and zero mean orbital polarization in a beyond band theory description of this system. 
In other words, the DFT+$U$ broken symmetry solution for  $Fm\bar{3}m$ is physically incorrect.

\begin{figure}
\begin{center}
\includegraphics[width=0.45\textwidth, angle=0]{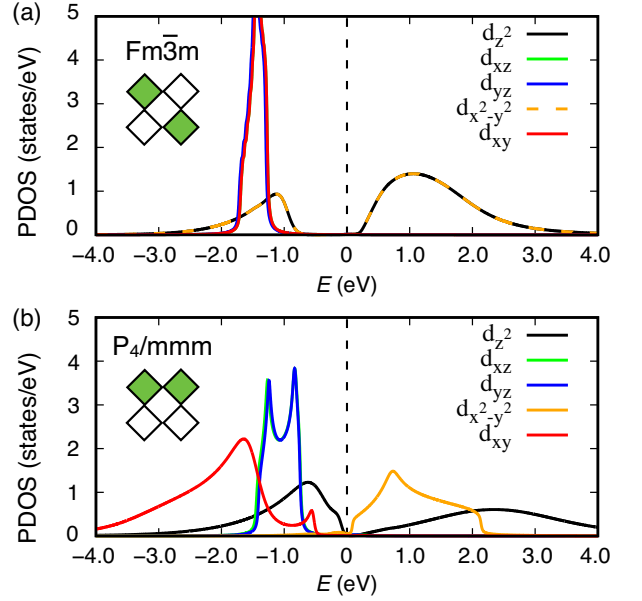}
\caption{DFT+DMFT spectral functions of Co $d$ Wannier orbitals for the low-spin state: (a) La$_2$CoTiO$_6$ ($Fm\bar{3}m$ structure) and (b) LTO+LCO superlattice ($P_{4}/mmm$, $a=b=c$ structure).  The calculations use $U=3$ eV,  $J=0.5$ eV, and a temperature of 300 K. 
}
\label{dmft_ls}
\end{center}
\end{figure}

\begin{table}
\begin{center}
\caption{Co $d$ Wannier occupancies  $N_d$  for the low-spin states within DFT+DMFT calculations. The calculations use $U=3$ eV,  $J=0.5$ eV, and a temperature of 300 K.}
\label{dmft-table-ls}
\renewcommand{\arraystretch}{1.5}
\begin{tabular}{c c c  c  c  c}
 \hline
	& $d_{z^2}$	& $d_{x^2-y^2}$	& $d_{xy}$	&$d_{xz}$	&	$d_{yz}$ \\
 \hline
(i) $Fm\bar{3}m$ La$_2$CoTiO$_6$	& 0.5		& 0.5		& 2.0 	& 2.0		& 2.0 \\	 
\hline
(ii) \makecell{$P_{4}/mmm$ LCO+LTO\\ ($a=b=c$)}	& 1.0		& 0.0		& 2.0		& 2.0		& 2.0 \\	 
\hline
\end{tabular}
\end{center}
\end{table}

We explicitly verify the artificial nature of the DFT+$U$ broken symmetry by performing DFT+DMFT calculations on the $Fm\bar{3}m$ phase with $a$=3.891 \AA\ in the paramagnetic phase. We use $U=3$ eV which reproduces the energy gap of bulk LaCoO$_3$ \cite{PRLKarolak2015}.  Orbitally resolved DFT+DMFT spectral functions of the Co $d$ Wannier orbitals in the LS $Fm\bar{3}m$ structure are presented in Fig. \ref{dmft_ls}.
We use $J=0.5$ eV obtain the LS Co$^{2+}$ state (the HS state 
is not stable with small $J$, much like Co$^{3+}$ in bulk LaCoO$_3$).
We show the DFT+DMFT electronic occupation $N_d$ of Co $d$ Wannier orbitals in Table \ref{dmft-table-ls}.
For the LS $Fm\bar{3}m$ phase of La$_2$CoTiO$_6$, Fig.~\ref{dmft_ls}(a) and  Table~\ref{dmft-table-ls}
 show clearly that the $e_{g}$ bands are degenerate and equally occupied due to  quantum fluctuations of electronic occupancy between $d_{z^2}$ and $d_{x^2-y^2}$. 
Therefore, the multi-determinant nature of the actual ground state washes out the $e_g$ orbital polarization of the single-determinant DFT+$U$ predictions.

\subsubsection{LS $P_{4}/mmm$ and $P2_{1}/n$ states}

Next, we consider the $P_{4}/mmm$ superlattice with $a=b=c$ (Figs. \ref{str_sym}(b) and \ref{orbpol_eg}(b)).
In this phase, the nearest neighbor environment of each Co is still perfectly cubic just as for the $Fm\bar{3}m$ phase, but the global cubic symmetry is broken by the formation of the superlattice along (001).  Therefore, the two $e_g$ bands are no longer degenerate even at $U_{\textnormal{Co}}=0$ (see the Appendix \ref{appendix:pdos} for relevant densities of states).  Clearly, $U_{\textnormal{Co}}\ne0$ is not a \emph{necessary} condition to split the $e_g$ degeneracy: as expected, symmetry reduction by forming a superlattice
is enough, but $U_{\textnormal{Co}}>0$ enhances the magnitude of $P$ substantially.
The orbital polarization is small but negative for  $U_{\textnormal{Co}}\leq 1$ eV but becomes substantially negative once  $U_{\textnormal{Co}}\geq2$ eV (i.e., $d_{x^2-y^2}$ is more occupied than $d_{z^2}$).  
Table ~\ref{energytable} shows that the  $d_{x^2+y^2,\uparrow}^1$ configuration is lower in energy than $d_{z^2,\uparrow}^1$ by 30 meV/Co
when $U_{\textnormal{Co}}=5$ eV: the orbital polarization should survive fluctuations and exist in the interacting realization. We note that for this system, $P(e_g)<0$ with DFT+$U$ for all $U_{\textnormal{Co}}$   considered.

We employ DFT+DMFT calcultions for the  $P_{4}/mmm$ superlattice with paramagnetic order to 
verify that the orbital polarization survives with quantum fluctuations.  
The associated spectral functions 
are presented in Fig.~\ref{dmft_ls}(b) and the Wannier occupacies in Table~\ref{dmft-table-ls}.
Since the structural symmetry is reduced, the electronic symmetry is broken, the $e_g$ degeneracy is split and the orbital polarization is evident, all of which is consistent with the DFT+$U$ results. 
Interestingly, the $d_{z^2}$ band is lower in energy than the  $d_{x^2-y^2}$ band within DFT+DMFT which disagrees with the DFT+$U$ result.
We expect that the difference might be due to the different $U$ dependence on the polarization 
for the Wannier $d$-only model, 
but this is a topic for future investigations.

We now move to the $P_{4}/mmm$ phase with $a=b\neq c$  (Figs. \ref{str_sym}(c) and \ref{orbpol_eg}(c)).
In this structure, the Co ions experience a tetragonal environment due to the relaxation. Similar to the previous $P_{4}/mmm$ $a=b=c$ case, the $e_g$ degeneracy is broken even at $U_{\textnormal{Co}}=0$, and the polarization magnitude is enhanced by $U_{\textnormal{Co}}>0$.  
Most notably, the $P$ for the $P_{4}/mmm$ ($a=b\ne c$) phase can be  negative or positive depending on the choices of  $U_{\textnormal{Co}}$ and $U_{\textnormal{Ti}}$ values
(see Fig. \ref{orbpol_eg}(c)). 
While it is clear that $U_{\textnormal{Co}}$ changes the splitting of Co $3d$ bands and also the magnitude of $P$, it is particularly interesting that $P$ also  depends strongly on $U_{\textnormal{Ti}}$ 
(compare the three $U_{\textnormal{Co}}=5$ eV results in Fig. \ref{orbpol_eg}(c)).
Since LCO+LTO is a charge-transfer heterostructure, $U_{\textnormal{Ti}}$ determines the amount of electron transfer from Ti to Co by adjusting the energy of the Ti $3d$ orbitals. Specifically, larger $U_{\textnormal{Ti}}$ results the higher energy $3d$ states and thus a larger amount of electron transfer to Co. 
Larger transfer induces stronger local electric fields from the TiO$_2$ to CoO$_2$ layers, and the field pushes the oxygen anions and increases out-of-plane Co--O bond lengths. The relation between the apical Co--O bond length and $P$ is explained by  simple crystal field theory.  Long out-of-plane Co--O bonds result the lowering of the energy of the out-of-plane orbital ($d_{z^2}$) since O is farther from Co along the $c$ axis, and thus $d_{z^2}$ becomes more occupied and  $P>0$.  Conversely, shorter out-of-plane Co--O bonds increase the energy of the $d_{z^2}$ band, so  $d_{x^2-y^2}$ becomes more occupied and $P<0$.   
We find that when $P<0$, the Co $d$ bands are always metallic.  On the other hand, when $P>0$ and large enough, the two $e_g$ bands are completely split in energy, and the system is in the insulating regime.

Finally, we consider the $P2_{1}/n$ phase which is our most stable structural phase.  Similar to the $P_{4}/mmm$ ($a=b=c$) and $P_{4}/mmm$ ($a=b\ne c$) phases, $P\ne0$ at $U_{\textnormal{Co}}=0$ and increases as a function of $U_{\textnormal{Co}}$.  As shown in Fig. \ref{orbpol_eg}(d), the $P$ of the $P2_{1}/n$ phase is always positive, as per our previous work \cite{part1}. The $d_{z^2}$ band is significantly lower in energy when $U_{\textnormal{Co}}=0$ and the material is insulating due to the energy  splitting in the $e_g$ manifold (see the Appendix \ref{appendix:pdos} for plots of the densities of states).

The sign of the orbital polarization $P$ is one of the interesting features of our results.
Since $P$ can be both positive and negative for the $P_{4}/mmm$ ($a=b\ne c$) phase, 
it is clear that the sign of $P$ is not due to the space group symmetry reduction alone.
Indeed, it is strongly determined by the local octahedral distortions, i.e., the relative   in-plane and out-of-plane Co--O bond lengths.
In the superlattice, the out-of-plane Co--O bond is well elongated by the local electric field between Co and Ti ions \cite{part1}.  Since the Co has interfaces at both sides and thus both of its out-of-plane Co--O bonds are elongated, the octahedral distortion of CoO$_6$ in (LCO)$_1$+(LTO)$_1$ has standardized symmetry label $A_{1g}+E_g$~\cite{PhysRevB.90.014308}. 
If the in-plane Co--O bond is longer than out-of-plane Co--O bond, $P$ becomes positive (Figs.~ \ref{orbpol_eg}(e) and (f)).
If the out-of-plane Co--O bond is longer than in-of-plane Co--O bond, $P$ becomes negative (Fig. \ref{orbpol_eg}(e)).  

\begin{figure}
\begin{center}
\includegraphics[width=0.35\textwidth, angle=0]{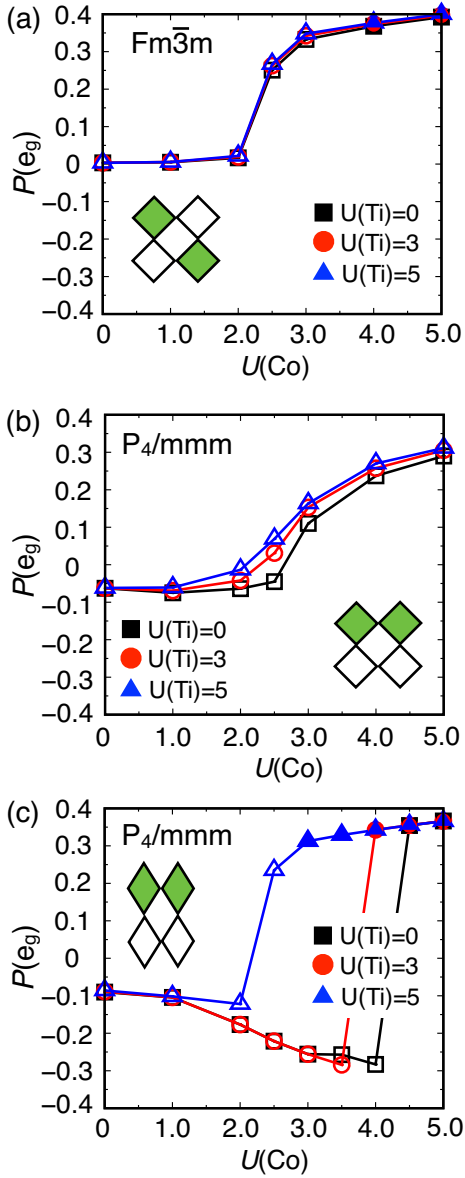}
\caption{  (a)-(c) show the DFT+$U$ orbital polarization $P(e_g)$ of the LS state structures with 
    different space groups versus $U({\rm Co})$ and as a function of $U({\rm Ti})$ 
    when the magnetic order between Co is antiferromagnetic (AFM).
Empty and filled points indicate metallic and insulating phases, respectively.
Insets show schematic side views of the octahedral tilts and distortions of the CoO$_6$ and TiO$_6$ oxygen octahedra. 
}
\label{AFM_orb_pol}
\end{center}
\end{figure}

\subsubsection{LS $e_g$ orbital polarization: magnetic interactions}
\label{sec:spin-spin}

Thus far, we have discussed the origin of the orbital polarization where we have imposed ferromagnetic (FM) ordering of the Co cations.  
As we will see, 
the effect of different magnetic ordering might change the sign of the polarization,
but it does not change any of our findings concerning the origin and magnitude  of the orbital polarization. 

To study the effect of the magnetic order on the polarization, we consider the antiferromagnetic (AFM) order for the $Fm\bar{3}m$ and $P_{4}/mmm$ ($a=b=c$) phases using supercells containing 2 distinct Co atoms and compare to the FM phase. The DFT+$U$ calculated orbital polarization is non-zero for the AFM $Fm\bar{3}m$ phase as shown in Fig.~\ref{AFM_orb_pol}: just like the FM phase, the DFT+$U$ energy is minimized by spontaneous symmetry breaking.  However, just like the FM phase, this is an artificial result, and quantum fluctuations should also wash out this orbital polarization.  The main reason is that the FM and AFM configurations are very close in energy, so that the actual system will be paramagnetic at any reasonable temperature.  Due to the large distance between Co atoms in the double perovskite structure (larger than 5.5 \AA)   as per Fig. \ref{str_sym}(a), the magnetic interaction between the Co cations is almost negligible: we find that the energies of the FM and AFM phases differ by only 0.3 meV/Co (for $U({\rm Ti})=5$ eV and $U({\rm Co})=5$ eV) .  Hence, the system is essentially paramagnetic, and our explicit DFT+DMFT calculations for the paramagnetic phase found no orbital polarization.

The overall behavior of the polarization of the $P_{4}/mmm$ structure with AFM ordering is also quite similar to the 
FM counterpart as shown in Figs.~\ref{AFM_orb_pol}(b) and (c).
Again, the structural symmetry reduction is the origin of the orbital polarization while $U$ enhances the polarization strongly.

\subsubsection{LS $e_g$ orbital polarization: strain dependence}
\label{sec:strain}

\begin{figure}
\begin{center}
\includegraphics[width=0.48\textwidth, angle=0]{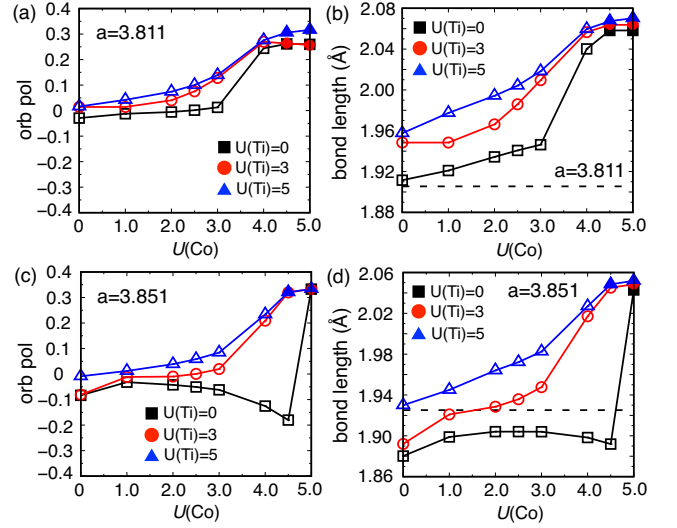}
\caption{  (a), (c) orbital polarization from DFT+$U$ of the LCO+LTO superlattice ($P_{4}/mmm$, $a^0a^0a^0$, and $a=b\neq c$) with different in-plane lattice parameters.
Empty and filled points indicate metallic and insulating phases, respectively. 
(b), (d) out-of-plane Co--O bond lengths with different in-plane lattice parameters.
Dashed lines represent the in-plane Co--O bond lengths, which weak functions of $U$.
}
\label{orbpol_strain}
\end{center}
\end{figure}

Now we discuss the effect of the strain on the $e_g$ orbital polarization for the LS phase.
Since the sign and the magnitude of  $P(e_g)$ depend on the relative sizes of the in-plane and out-of-plane Co--O bonds,  strain can enhance, reduce, or change the sign of $P$, since the Co--O bond lengths can be strongly altered by the epitaxial strain. We consider the $P_{4}/mmm$ ($a=b\neq c$) phase with in-plane lattice parameters of $a=3.811$ and  3.851 \AA\, and the results are summarized in Fig.~\ref{orbpol_strain} (the data for $a=3.891$ \AA\ is in Figs. \ref{orbpol_eg}(c) and (e)).

For $a=3.811$ \AA, where the CoO$_6$ octahedra feel compressive strain, 
apical Co--O bonds are always longer than the in-plane Co--O bonds.  Thus, the $d_{z^2}$ band is always lower in energy than the $d_{x^2-y^2}$ band, and $P>0$ as per simple crystal field theory. 
In addition,  $U_{\textnormal{Co}}>0$ further increases the splitting between the $e_g$ bands;
as a result, both apical Co--O bond lengths and $P$ are monotonically increasing functions of $U_{\textnormal{Co}}$.

For $a=3.851$ \AA, which represents weaker compressive strain, the apical bonds are elongated but 
not always longer than the in-plane bonds. 
Therefore, similar to the $a=3.891$ \AA\ case, the sign of $P$ depends on both 
$U_{\textnormal{Co}}$ and $U_{\textnormal{Ti}}$.
The biggest difference between $a=3.851$  and 3.891 \AA\ is evident for the $(U_{\textnormal{Co}}=5$, $U_{\textnormal{Ti}}=0)$ case: 
 $P>0$ for $a=3.851$ \AA\ but $P<0$ for $a=3.891$ \AA.

\subsection{HS state: $t_{2g}$ orbital polarization}
\label{sec:t2g}

\begin{figure}
\begin{center}
\includegraphics[width=0.48\textwidth, angle=0]{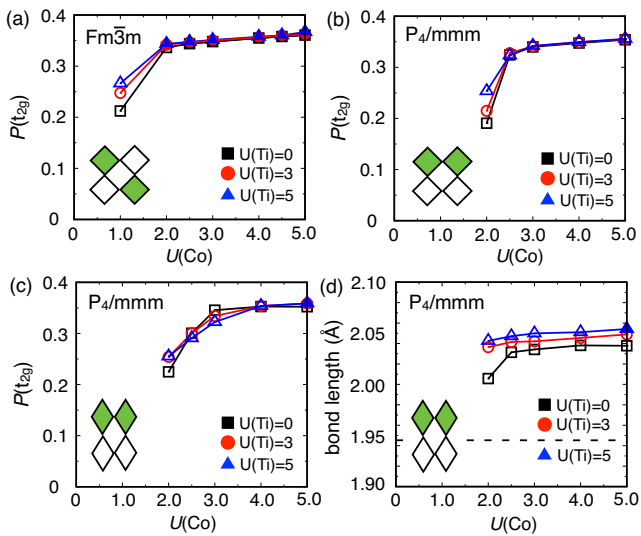}
\caption{  (a)-(c) $t_{2g}$ orbital polarization from DFT+$U$ of the LCO+LTO heterostructures with different space groups; as shown by the insets, (b) has $a=b=c$ and (c) has $a=b\neq c$. Filled and empty points indicate metallic and insulating phases, respectively.  
(d) out-of-plane Co--O bond lengths of the $P_{4}/mmm$ phase with $a=b\neq c$.
The dashed line represents the in-plane Co--O bond lengths, which are robust versus $U$.
 }
\label{orbpol_t2g}
\end{center}
\end{figure}

In this section, we consider the orbital polarization of the $t_{2g}$ bands, the relevant quantity for the HS Co$^{2+}$ spin state.
In prior work, we have shown that the orbital ordering is enhanced for the 
HS $P2_{1}/n$ phase by large tensile strain \cite{part1}.
However, for the range of in-plane lattice parameters considered here, the orbital ordering is almost negligible: the magnitude of the JT distortion (difference between two in-plane Co--O bonds) is less than 0.007 \AA. For what follows below, we average the orbital polarization of two Co atoms, but we note that this averaged value and the values from either Co atom are almost identical. 

 We define the orbital polarization of the high-spin (HS) state   as 
\begin{equation}
P(t_{2g}) = \frac{(n^{\uparrow}_{xz}+n^{\downarrow}_{xz})+ (n^{\uparrow}_{yz}+n^{\downarrow}_{yz}) - 2(n^{\uparrow}_{xy}+n^{\downarrow}_{xy})  }
{(n^{\uparrow}_{xz}+n^{\downarrow}_{xz})+ (n^{\uparrow}_{yz}+n^{\downarrow}_{yz})+ 2(n^{\uparrow}_{xy}+n^{\downarrow}_{xy})  }\,.
\end{equation}

We consider three structures: (i) $Fm\bar{3}m$ space group and $a^0a^0a^0$ tilt, (ii) $P_{4}/mmm$ with $a=b=c$, (iii) and $P_{4}/mmm$ with $a=b\neq c$.  
We do not examine the $P2_{1}/n$ case: the local $t_{2g}$ states on each Co become mixed due to the octahedral tilts, and the off-diagonal elements of the density matrix in the $t_{2g}$ manifold become large and non-negligible: this makes unambiguous extraction of individual orbital occupancies difficult.

Fig.~\ref{orbpol_t2g}(a) shows that $P$ for the highest symmetry $Fm\bar 3 m$ structure is generally non-zero for even modest $U_{\rm Co}$ values: this means that the $t_{2g}$ subsystem has a stronger propensity to spontaneously break electronic symmetry at the DFT+$U$ level when compared to the $e_g$ system above.  We believe this is due to the narrower $t_{2g}$ energy bands and the more localized electronic states on the Co cations.  However, the total energies of the three equivalent configurations $d_{xy\downarrow}^1 d_{xz\downarrow}^1 d_{yz\downarrow}^0$, 
$d_{xy\downarrow}^1 d_{xz\downarrow}^0 d_{yz\downarrow}^1$,
and $d_{xy\downarrow}^0 d_{xz\downarrow}^1 d_{yz\downarrow}^1$ differ by only 0.1 meV/Co (see Table~\ref{energytable}).  Again, this indicates that the actual interacting $Fm\bar 3 m$  system should have significant fluctuations between these configurations and zero mean orbital polarization.

Next, in both $P_{4}/mmm$ phases, we expect the orbital polarization predicted in Figs.~\ref{orbpol_t2g}(b,c) to be observable because, as Table~\ref{energytable} shows, the $d_{xy\downarrow}^0 d_{xz\downarrow}^1 d_{yz\downarrow}^1$ configuration has significantly lower energy than the other competing configurations (which are the orbitally ordered $d_{xy\downarrow}^1 d_{xz\downarrow}^1 d_{yz\downarrow}^0$ and $d_{xy\downarrow}^1 d_{xz\downarrow}^0 d_{yz\downarrow}^1$ systems).

Permitting the local octahedra to elongate in going from the $P_{4}/mmm$ $a=b=c$ to the $a=b\ne c$ phase (Fig.~\ref{orbpol_t2g}(b) to Fig.~\ref{orbpol_t2g}(c)) increases the polarization $P$.  The main difference from the $e_g$ case  is that the sign of $P$ is insensitive to the value of both $U_{\rm Co}$ and $U_{\rm Ti}$.  This goes hand in hand with the structure of the system: Fig.~\ref{orbpol_t2g}(d) shows that the HS $t_{2g}$ system has longer out-of-plane Co--O bonds than the $e_g$ LS case, and its out-of-plane bonds are always longer than the in-plane bonds.

\begin{figure}
\begin{center}
\includegraphics[width=0.45\textwidth, angle=0]{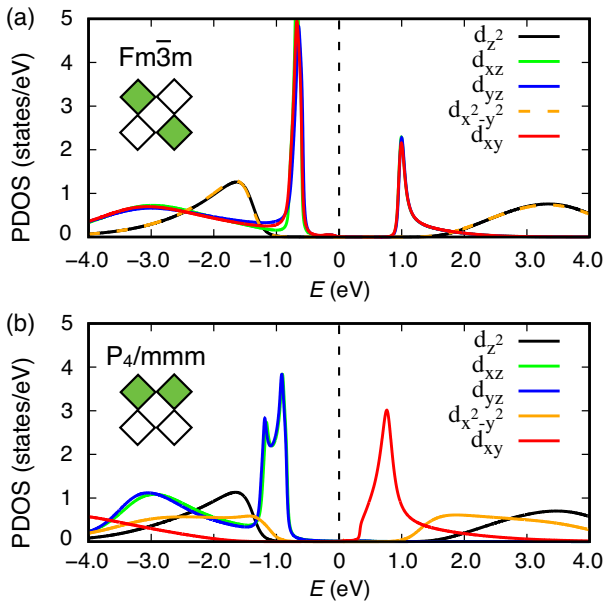}
\caption{DFT+DMFT spectral functions of Co $d$ Wannier orbitals for (a) HS La$_2$CoTiO$_6$ ($Fm\bar{3}m$) and (b) the HS LTO+LCO superlattice ($P_{4}/mmm$, $a=b=c$).
The calculations use $U=3$ eV,  $J=0.9$ eV, and a temperature of 300 K. 
}
\label{dmft_hs}
\end{center}
\end{figure}

\begin{table}
\begin{center}
\caption{Wannier Co $d$ occupancies $N_d$ for the high-spin states within DFT+DMFT calculations. The calculations use $U=3$ eV,  $J=0.9$ eV, and a temperature of 300 K.}
\label{dmft-table-hs}
\renewcommand{\arraystretch}{1.5}
\begin{tabular}{c c c  c  c  c}
 \hline
	& $d_{z^2}$	& $d_{x^2-y^2}$	& $d_{xy}$	&$d_{xz}$	&	$d_{yz}$ \\
 \hline
(i) $Fm\bar{3}m$ La$_2$CoTiO$_6$	& 1.00	& 1.00	& 1.67	& 1.67	& 1.67 \\	 
\hline
(ii) \makecell{$P_{4}/mmm$ LCO+LTO \\ ($a=b=c$)}	& 1.00	& 1.00	& 1.00	& 2.00	& 2.00 \\	 
\hline
\end{tabular}
\end{center}
\end{table}

To study the importance of quantum fluctuations, we repeat our procedure for the LS phases.  We employ DFT+DMFT calculations for the HS states of La$_2$CoTiO$_6$ ($Fm\bar{3}m$ structure) and LTO+LCO superlattice ($P_{4}/mmm$, $a=b=c$ structure), and the results are summarized in Fig. \ref{dmft_hs} and Table \ref{dmft-table-hs} (we use a larger value of $J=0.9$ eV to stabilize the HS spin configuration).
Much like the LS case, the HS $Fm\bar{3}m$ phase retains its $t_{2g}$ degeneracy within DFT+DMFT due to the quantum fluctuation between the $t_{2g}$ orbitals and shows no orbital polarization.  On the other hand, for the HS $P_{4}/mmm$ phase, the $t_{2g}$ bands split into doubly-degenerate 
$d_{xz}$+$d_{yz}$ bands and non-degenerate $d_{xy}$ band as shown in Fig. \ref{dmft_hs}(b): the superlatticing effect along is sufficient to generate and stabilize the orbital polarization.

\section{Summary}

In this work, we have shown that orbital polarization in the cobaltate systems we have studied has its fundamental origin in the structure and symmetry of the material (the crystalline environment of the Co cations); strong electronic interactions can enhance the polarization but are not necessary to generate it.  The crystalline symmetry reduction in our cases is oxygen octahedral elongation which then breaks the electronic symmetry in the Co $e_g$ and $t_{2g}$ manifolds.  

We note that this symmetry breaking mode is simple and in some sense crude: it breaks symmetry in both the $e_g$ and $t_{2g}$ and is unable to selectively do so in one or the other manifold.  In principle, we can envision symmetry breaking operations that selectively remove degeneracy in only one manifold, but they require control over the electronic potential in a fine-grained microscopic manner that goes beyond simply distorting cation-oxygen bonds.  Whether such an advanced level of control is feasible in actual materials is, in our mind, an interesting open question.

\section{Acknowledgments}
We thank to Sangjae Lee, F. J. Walker, and Charles H. Ahn for helpful discussions. We thank the Yale Center for Research Computing for guidance and use of the research computing infrastructure.  This work also used the Extreme Science and Engineering Discovery Environment (XSEDE), which is supported by National Science Foundation grant number ACI-1548562, by using computer time on the Comet supercomputer as enabled by XSEDE allocation MCA08X007.

\bibliography{myref}

\section{Apendix A: Effect of SOC}
\label{appendix:soc}

\begin{table*}
\begin{ruledtabular}
\begin{center}
\caption{Eigenvalues ($\nu_i$, last column) and eigenvector components squared (probabilities) of the 10$\times$10 single-particle density matrix of the Co $3d$ manifold
obtained within GGA+$U$+SOC calculations. 
Structural phase is $P_{4}/mmm$ ($a=b\neq c$) with $U_{\rm Ti}$=5 eV and $U_{\rm Co}$=5 eV and LS cobalt spin state.  Each row describes one eigenvalue/eigenvector.
}
\label{soctable}
\renewcommand{\arraystretch}{1.3}
\begin{tabular}{c | c c c c c   c c c c c|c}
  &$d^{\uparrow}_{xy}$   &$d^{\uparrow}_{yz}$  &$d^{\uparrow}_{z^2}$  &$d^{\uparrow}_{xz}$  &$d^{\uparrow}_{x^2-y^2}$  
  &$d^{\downarrow}_{xy}$  &$d^{\downarrow}_{yz}$  &$d^{\downarrow}_{z^2}$  &$d^{\downarrow}_{xz}$  &$d^{\downarrow}_{x^2-y^2}$ 
& $\nu_i$  \\
\hline
$\nu_1$	&0.000	&0.046	&0.000	&0.046	&0.000	&0.002	&0.000	&0.000	&0.000	&0.906	&0.146  \\
$\nu_2$	&0.000	&0.000	&0.000	&0.000	&0.000	&0.000	&0.000	&1.000	&0.000	&0.000	&0.156  \\
$\nu_3$	&0.003	&0.000	&0.000	&0.000	&0.997	&0.000	&0.000	&0.000	&0.000	&0.000	&0.393   \\
$\nu_4$	&0.000	&0.000	&0.053	&0.000	&0.000	&0.000	&0.473	&0.000	&0.473	&0.000	&0.865  \\
$\nu_5$	&0.002	&0.000	&0.000	&0.000	&0.000	&0.000	&0.499	&0.000	&0.499	&0.000	&0.947  \\
$\nu_6$	&0.000	&0.040	&0.000	&0.040	&0.000	&0.918	&0.000	&0.000	&0.000	&0.002	&0.952  \\
$\nu_7$	&0.000	&0.500	&0.000	&0.500	&0.000	&0.000	&0.000	&0.000	&0.000	&0.000	&0.958  \\
$\nu_8$	&0.995	&0.000	&0.000	&0.000	&0.003	&0.000	&0.001	&0.000	&0.001	&0.000	&0.968  \\
$\nu_9$	&0.000	&0.000	&0.947	&0.000	&0.000	&0.000	&0.027	&0.000	&0.027	&0.000	&1.023  \\
$\nu_{10}$	&0.000	&0.414	&0.000	&0.414	&0.000	&0.079	&0.000	&0.000	&0.000	&0.092 	&1.047  \\
\end{tabular}
\end{center}
\end{ruledtabular}
\end{table*}

It is known that the strength of the spin-orbit coupling (SOC)  is small for transition metal 3$d$ orbitals \cite{Cole1970}.  For Co,  experiments show that the SOC constant of Co$^{2+}$ is 16 meV \cite{CowleyPRB2013}.
Given that the SOC is weak for Co$^{2+}$, we expect a very weak SOC effect on the orbital polarization, 
and thus we check only few cases using GGA+$U$+SOC calculations. Since SOC will break the block diagonal structure of the single-particle density matrix in the spin sector,  we use the eigenvalues of the single-particle density matrix of the entire Co $3d$ manifold.

For $e_g$ polarization, we choose 4 eigenvectors where the largest portion is $d^{\uparrow}_{z^2}$, $d^{\downarrow}_{z^2}$, 
$d^{\uparrow}_{x^2-y^2}$, and $d^{\downarrow}_{x^2-y^2}$, respectively, and use their eigenvalues 
to calculate the polarization.
For example, for the $P_{4}/mmm$ ($a=b\neq c$) LS phase with $U_{\rm Ti}$=5 eV and $U_{\rm Co}$=5 eV,
the polarization within SOC $P_{soc}(e_{g})$ is obtained by
\begin{equation}\label{eq:soc}
P_{soc}(e_{g})=\frac{ (\nu_9+\nu_2) -(\nu_3+\nu_1) }{(\nu_9+\nu_2) +(\nu_3+\nu_1)} \ ,
\end{equation}
where $\nu_i$ are related eigenvalues among the $10\times 10$ single-particle density matrix for the case of $d$ electrons,
shown in Table \ref{soctable}.
Not surprisingly given the  weak SOC strength for 3$d$ orbital, 
in this case the polarizations with and without SOC are $P(e_{g})=0.349$ and $P_{soc}(e_{g})=0.344$, respectively, which differ by $\sim$1\%.

\section{Appendix B: PDOS}
\label{appendix:pdos}
The figures in this appendix provide the projected densities of electronic states for many of the systems described in the main text.

\begin{figure}
\begin{center}
\includegraphics[width=0.48\textwidth, angle=0]{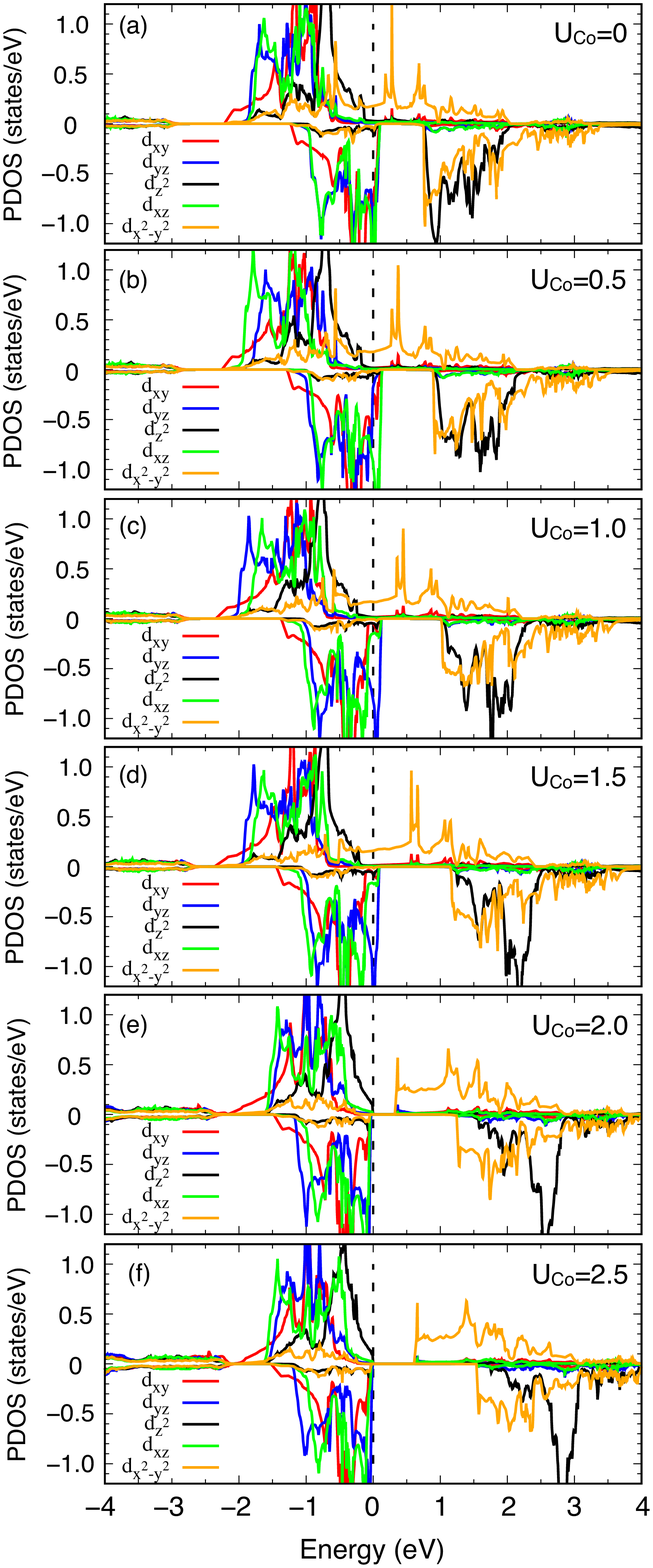}
\caption{      (a)-(f) Co 3$d$ projected density of states for low-spin (LS) Co in the  (LaCoO$_3$)$_1$+(LaTiO$_3$)$_1$ 
     superlattice as a function of  $U_{\textnormal{Co}}$ in eV. The value of $U_{\textnormal{Ti}}$ is 5 eV, 
     the in-plane lattice parameters $a$ and $b$ are fixed to 3.811 \AA, 
     and the atomic structure has the $P2_{1}/n$ space group.
}
\label{LS_Udep}
\end{center}
\end{figure}

\begin{figure}
\begin{center}
\includegraphics[width=0.48\textwidth, angle=0]{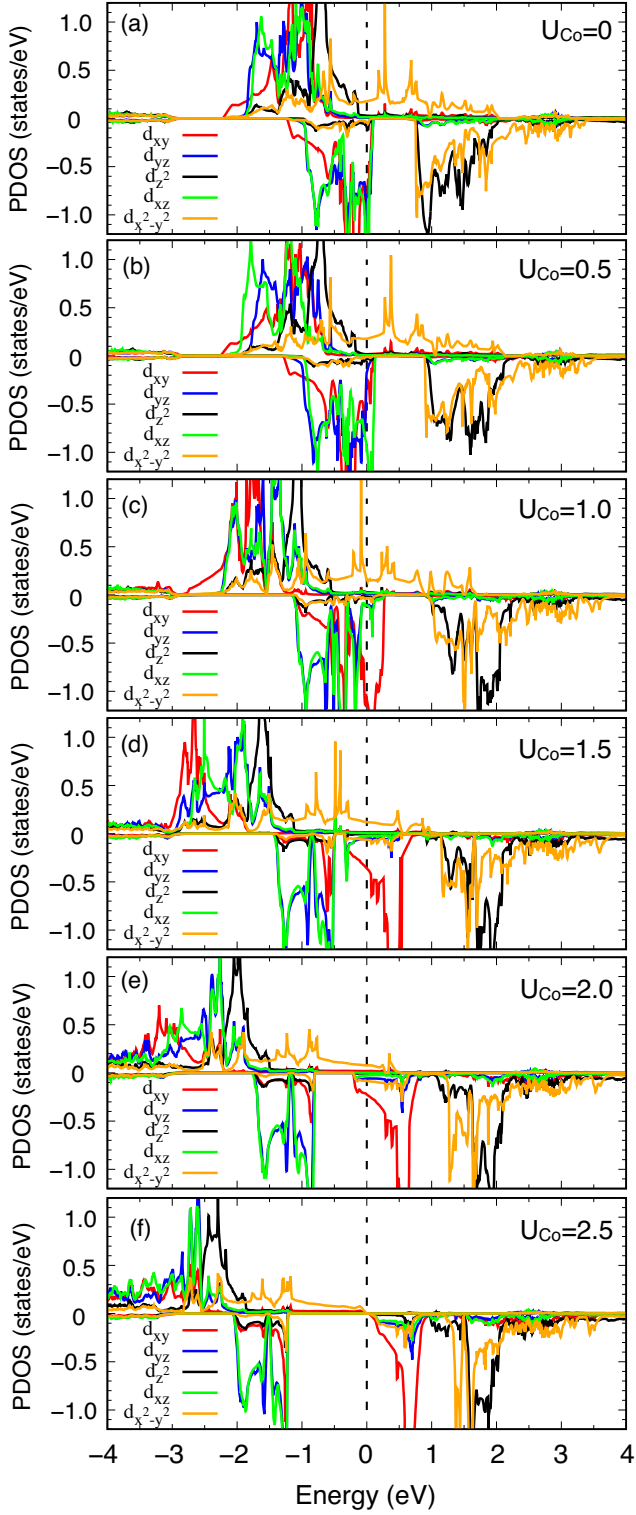}
\caption{    (a)-(f) Co 3$d$ projected density of states of high-spin (HS) Co in the  (LaCoO$_3$)$_1$+(LaTiO$_3$)$_1$ 
      superlattice as a function of $U_{\textnormal{Co}}$ in eV.
      The value of $U_{\textnormal{Ti}}$ is 5 eV, the in-plane lattice parameters $a$ and $b$ are fixed to 3.811 \AA, 
      and the atomic structure has the $P2_{1}/n$ space group.
     Note that for $U_{\textnormal{Co}}=$0 and 0.5, the HS state is not even metastable, 
     so the Co has the LS state. The in-plane lattice parameters $a$ and $b$ are fixed to 3.811 \AA.
}
\label{HS_Udep}
\end{center}
\end{figure}


\begin{figure}
\begin{center}
\includegraphics[width=0.48\textwidth, angle=0]{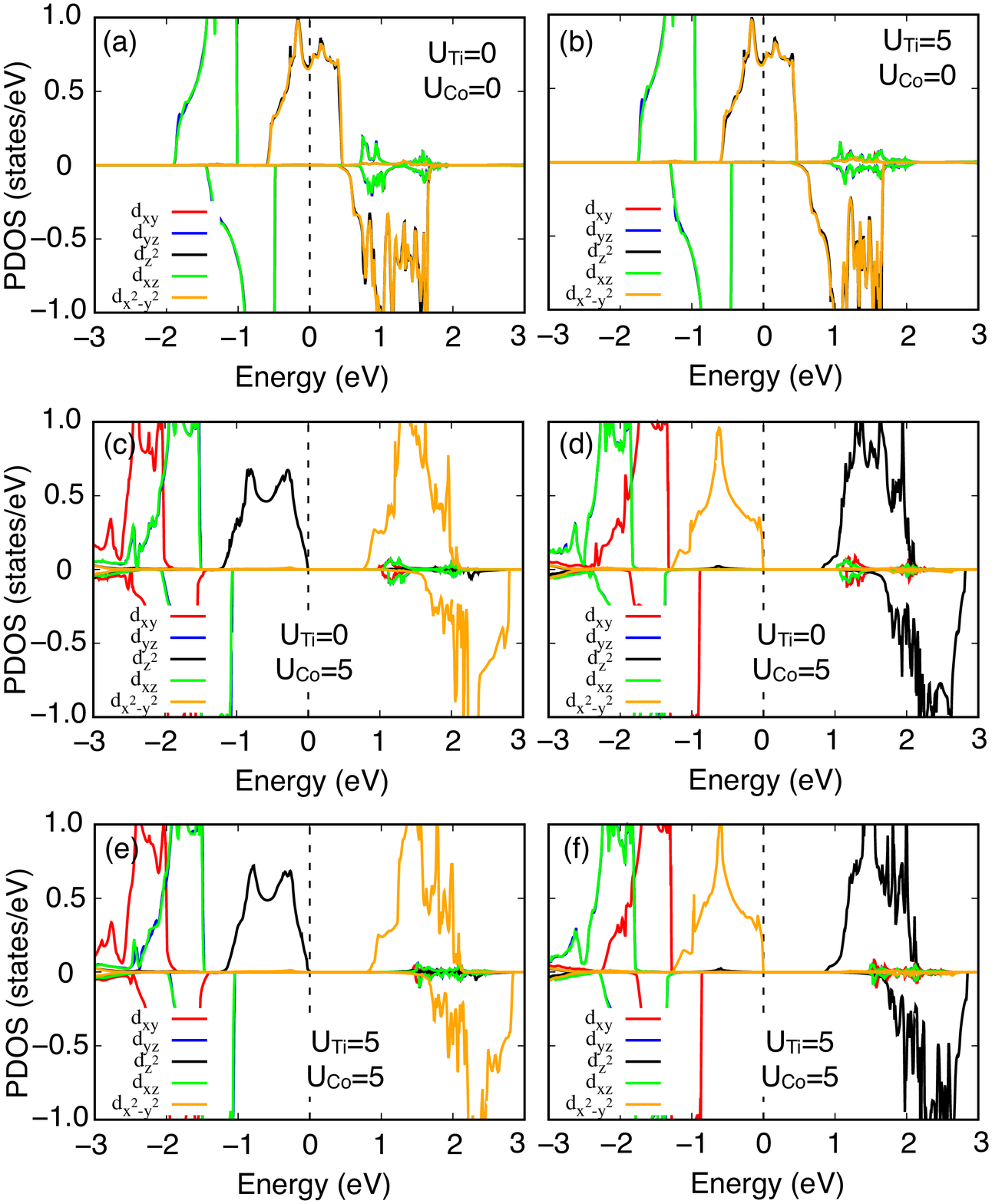}
\caption{  Comparison of the Co 3d projected density of states of the two different self-consistent solutions for low-spin (LS) Co in the double perovskite La$_2$TiCoO$_6$ (cubic, $Fm\bar{3}m$ space group) as a function of  $U_{\textnormal{Co}}$ and $U_{\textnormal{Ti}}$ in eV.
(c) and (e) are for the $d_{z^2}$-occupied LS states, and (d) and (f) are for the $d_{x^2-y^2}$-occupied LS states.
Lattice parameters are fixed to $a=b=c=3.891$ \AA, obtained by minimizing the stress of La$_2$TiCoO$_6$ with $U_{\textnormal{Co}}=U_{\textnormal{Ti}}=3$.
}
\label{eg_cubicDP_3.891}
\end{center}
\end{figure}

\begin{figure}
\begin{center}
\includegraphics[width=0.95\columnwidth, angle=0]{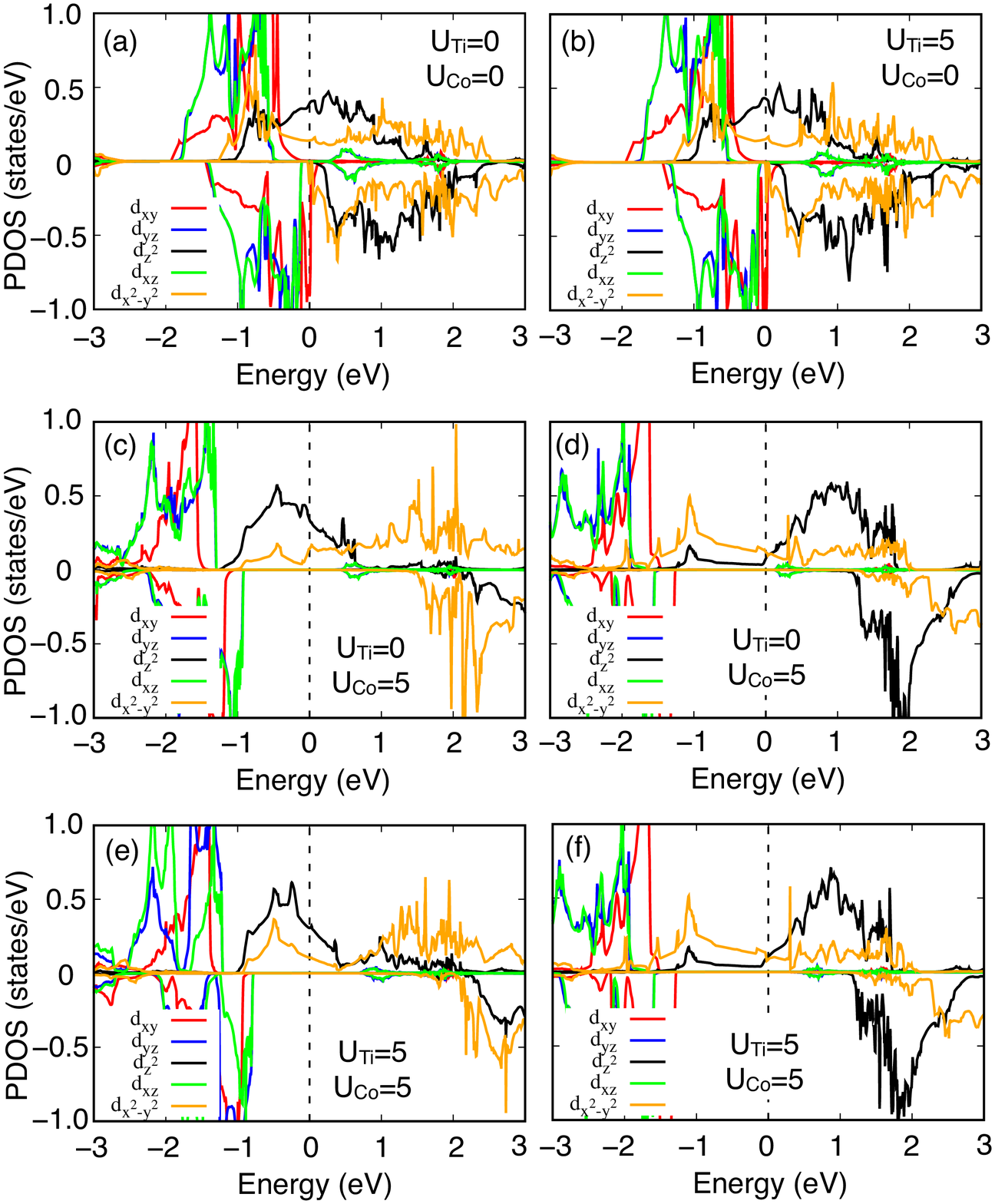}
\caption{Comparison of the Co 3d projected density of states of the two different self-consistent solutions for low-spin (LS) Co in the  (LaCoO$_3$)$_1$+(LaTiO$_3$)$_1$ superlattice 
($P_{4}/mmm$ space group, $a=b=c$) as a function of  $U_{\textnormal{Co}}$ and $U_{\textnormal{Ti}}$ in eV. 
(c) and (e) are for the $d_{z^2}$-occupied LS states, and (d) and, (f) are for the $d_{x^2-y^2}$-occupied LS states.
Lattice parameters are fixed to $a=b=c=3.891$ \AA.
}
\label{eg_cubicSL_3.891}
\end{center}
\end{figure}

\begin{figure}
\begin{center}
\includegraphics[width=0.95\columnwidth, angle=0]{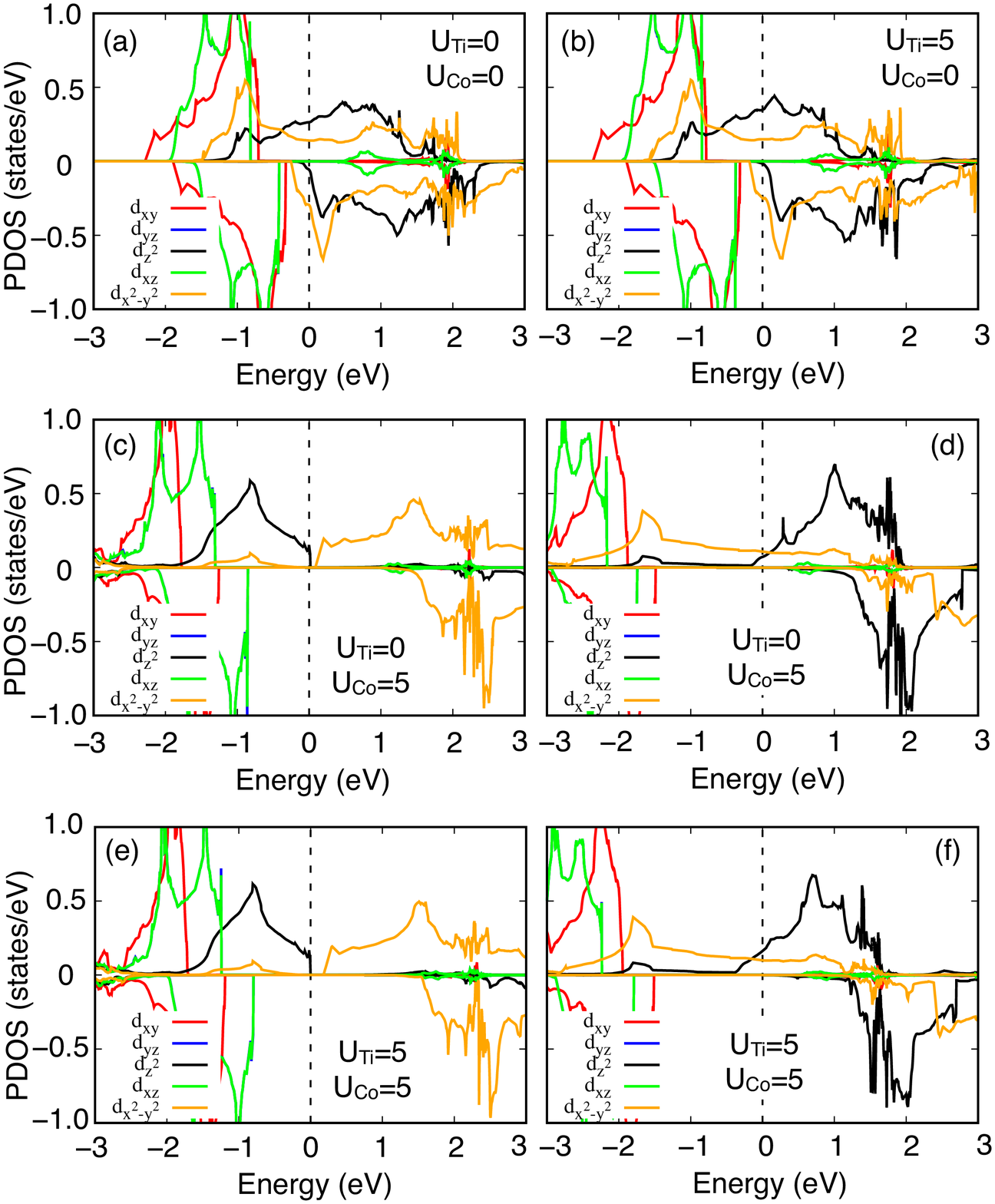}
\caption{Comparison of the Co 3d projected density of states of the two different self-consistent solutions for low-spin (LS) Co in (LaCoO$_3$)$_1$+(LaTiO$_3$)$_1$ 
($P_{4}/mmm$ space group, $a=b\neq c$) as a function of  $U_{\textnormal{Co}}$ and $U_{\textnormal{Ti}}$ in eV. 
(c) and (e) are for the $d_{z^2}$-occupied LS states, and (d) and (f) are for the $d_{x^2-y^2}$-occupied LS states.
In-plane lattice parameters are fixed to $a=b=3.891$ \AA, while $c$ is different due to the relaxation.
}
\label{eg_relaxZ_3.891}
\end{center}
\end{figure}

\begin{figure}
\begin{center}
\includegraphics[width=0.95\columnwidth, angle=0]{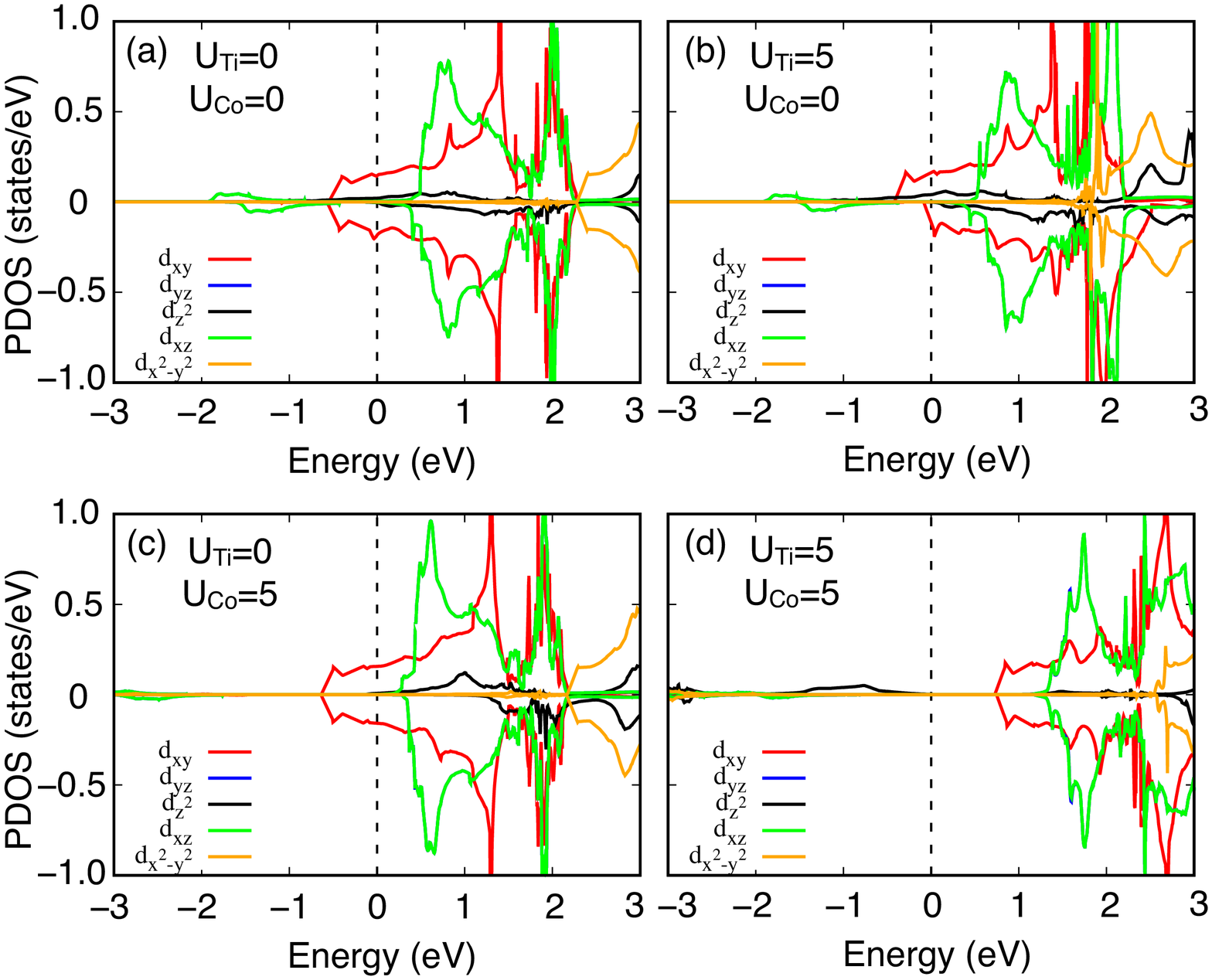}
\caption{Ti 3d projected density of states for the low-spin (LS) Co in the  (LaCoO$_3$)$_1$+(LaTiO$_3$)$_1$ superlattice
($P_{4}/mmm$ space group, $a=b\neq c$) as a function of  $U_{\textnormal{Co}}$ and $U_{\textnormal{Ti}}$ in eV. 
In-plane lattice parameters are fixed to $a=b=3.891$ \AA, while $c$ is different due to the relaxation.
}
\label{eg_relaxZ_3.891_Ti}
\end{center}
\end{figure}

\begin{figure}
\begin{center}
\includegraphics[width=0.95\columnwidth, angle=0]{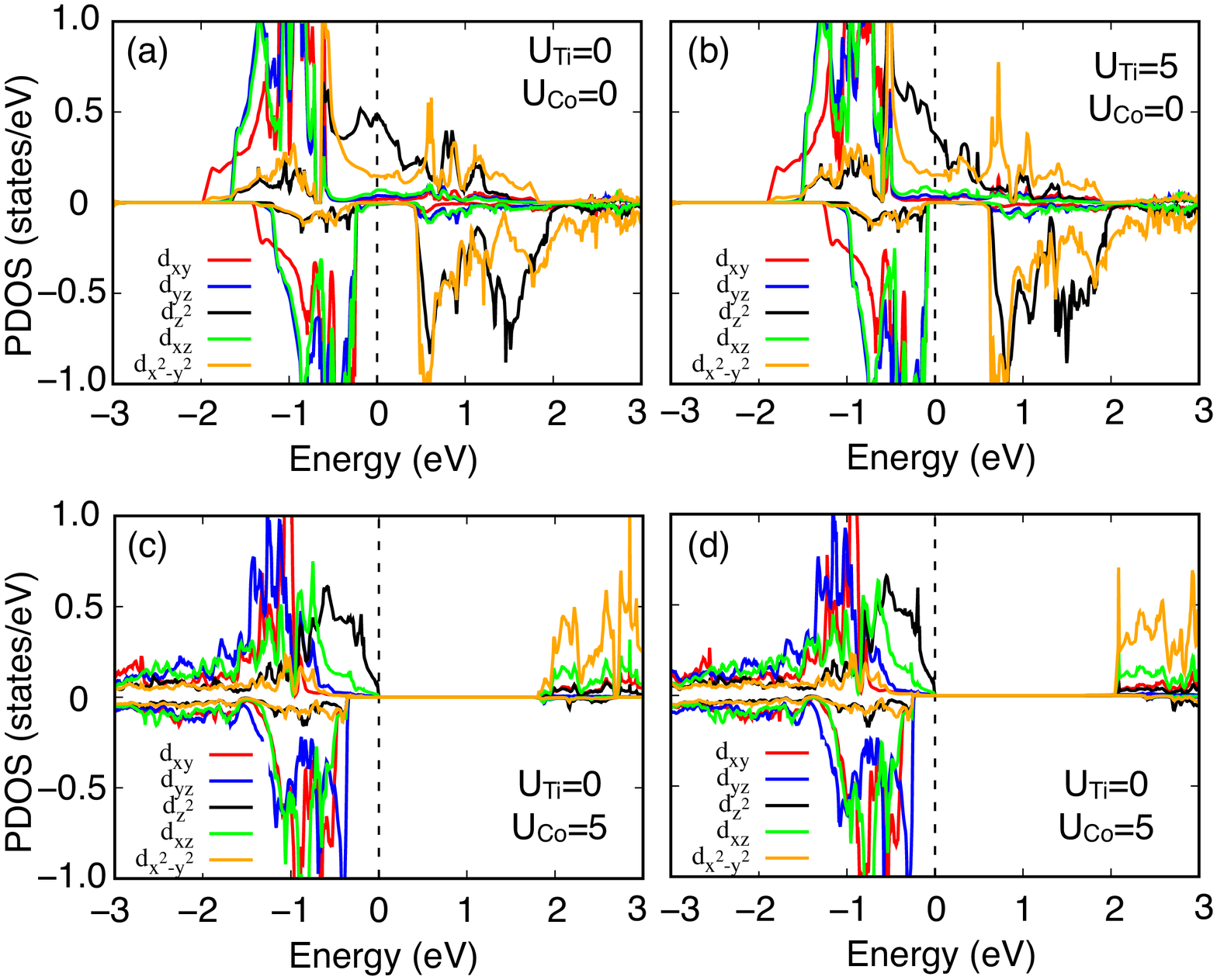}
\caption{Co 3d projected density of states for the low-spin (LS) Co in the  (LaCoO$_3$)$_1$+(LaTiO$_3$)$_1$ superlattice
($P2_{1}/n$ space group, $a=b\neq c$) as a function of  $U_{\textnormal{Co}}$ and $U_{\textnormal{Ti}}$ in eV.
In-plane lattice parameters are fixed to $a=b=3.891$ \AA, while $c$ is different due to the relaxation.
}
\label{eg_tilt_3.891}
\end{center}
\end{figure}


\begin{figure}
\begin{center}
\includegraphics[width=0.95\columnwidth, angle=0]{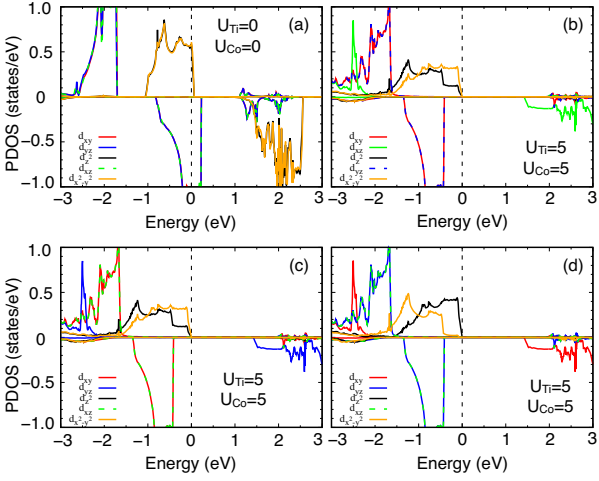}
\caption{Co 3d projected density of states (PDOS) for high-spin (HS) Co in the double perovskite La$_2$TiCoO$_6$ 
(cubic, $Fm\bar{3}m$ space group) as a function of  $U_{\textnormal{Co}}$ and $U_{\textnormal{Ti}}$ in eV. (a) For $U_{\textnormal{Ti}}=U_{\textnormal{Co}}=0$, the minority (down) $t_{2g}$ states are equally occupied showing no orbital polarization.  For $U_{\textnormal{Ti}}=U_{\textnormal{Co}}=5$ eV, three physically equivalent different minority $t_{2g}$ configurations can be stabilized: (b) $d_{xy\downarrow}^1 d_{xz\downarrow}^0 d_{yz\downarrow}^1$, 
(c) $d_{xy\downarrow}^1 d_{xz\downarrow}^1 d_{yz\downarrow}^0$, and 
(d) $d_{xy\downarrow}^0 d_{xz\downarrow}^1 d_{yz\downarrow}^1$.
Lattice parameters are fixed to $a=b=c=3.891$ \AA. }
\label{t2g_cubicDP_3.891}
\end{center}
\end{figure}

\end{document}